\documentclass[12pt]{iopart}

\usepackage{iopams}
\usepackage[english]{babel}
\usepackage{graphicx}
\usepackage{cite}
\usepackage{bbm}

\begin{document}

%


\title{Quantum harmonic oscillator with time dependent mass}

\author{I. Ramos-Prieto,$^{1}$ A. Espinosa-Zu\~niga,$^{1}$ M. Fern\'andez-Guasti,$^{2}$
and H.M. Moya-Cessa$^{1}$}

\address{$^{1}$INAOE, Coordinación de Optica, Apdo. Postal 51 y 216, 72000
Puebla, Pue., Mexico. \\
 $^{2}$Depto. de Física, CBI, Universidad Autónoma Metropolitana
- Iztapalapa, 09340, CDMX, Ap. Postal 55-534, Mexico. }

\begin{abstract}
We use the Fourier operator to transform a time dependent mass
quantum harmonic oscillator into a frequency dependent one. Then
we use Lewis-Ermakov invariants to solve the Schr\"odinger
equation by using squeeze operators. Finally we give two examples
of time dependencies: quadratically and hyperbolically growing
masses.
\end{abstract}

\pacs{42.50.Ct, 42.50.-p, 42.50.Pq, 42.50.Dv}


\maketitle

\section{Introduction}

The harmonic oscillator equation with time dependent parameters \cite{Lewis,Lewis2,Dodonov79,Ray,Janszky86,Pedrosa,Haas01,Bouquet96}
has been solved for a sudden frequency change using a continuous treatment
based on an invariant formalism \cite{Moya03}. This analytic treatment
requires that the time dependent parameter is a monotonic function
whose variation is short compared with the typical period of the system.
This procedure allowed us to obtain analytic solutions that do or
do not exhibit squeezing \cite{Yuen,Caves,Satya,Vidiella,Knight}
depending on time when departing from an initial coherent state.

Ion laser interactions \cite{Schleich} is one of the fields where
time dependent harmonic oscillator arise and some solutions have
been obtained \cite{Manko2,Schrade} for the time dependent
ion-laser Hamiltonian \cite{Segundo}.

On the other hand, time dependent mass harmonic oscillators have been
considered, probably the most famous of them being the so-called Caldirola-Kanai
oscillator \cite{Caldirola,Kanai} that has been studied extensively
over the years \cite{Tava1,Tava2,Kim,Manko}. A time dependent exponentially
growing mass is considered in this oscillator, while the stiffness
is also allowed to change in time. We have considered recently a time
dependent mass subject to a sudden change \cite{Guasti2} while keeping
the stiffness parameter constant.

{Propagation in one dimensional
optical metamaterials with graded refractive index can be efficiently
treated with the time dependent harmonic oscillator formalism. The
time axis is replaced by the wave direction of propagation and the
medium inhomogeneity is represented by the time dependent parameter.
Optical metamaterials are relevant for photonic crystals and other
state of the art optical applications \cite{Habib,Shvartsburg}. }

{Mass reconfiguration is being
used to manipulate oscillations \cite{Zevin,Wright} and chaotic behaviour
in macroscopic oscillators \cite{Stilling}. }

\section{Time dependent mass}

The harmonic Hamiltonian with time dependent mass $M(t)$ reads
\begin{equation}
\hat{H}(t)=\frac{1}{2}\left[\frac{\hat{p}^{2}}{M(t)}+\kappa\hat{q}^{2}\right]\label{td}
\end{equation}
where $\kappa$ is the stiffness. We want to solve the Schr\"odinger
equation
\begin{equation}
i\frac{\partial|\psi(t)\rangle}{\partial t}=\hat{H}(t)|\psi(t)\rangle\label{schr0}
\end{equation}
where we have set $\hbar=1$. We consider the Fourier operator \cite{Namias,Agarwal,Fan,Weimann}
\begin{equation}
\hat{{\mathcal{F}}}=\exp\left(-i\frac{\pi}{4}[\hat{p}^{2}+\hat{q}^{2}]\right)e^{i\pi/4}
\end{equation}
and do the transformation $|\psi(t)\rangle=\hat{\mathcal{F}}|\Psi(t)\rangle$.
Inserting this expression in (\ref{schr})
\begin{equation}
i\frac{\partial|\Psi(t)\rangle}{\partial t}=\frac{1}{2}\hat{\mathcal{F}}^{\dagger}\left[\frac{\hat{p}^{2}}{M(t)}+\kappa\hat{q}^{2}\right]\hat{\mathcal{F}}|\Psi(t)\rangle\label{schr}
\end{equation}
by rescaling time with the stiffness constant, $\tau=\kappa t$ we
rewrite the Schr\"odinger equation in the form
\begin{equation}
i\frac{\partial|\Psi(\tau)\rangle}{\partial\tau}=\frac{1}{2}\hat{\mathcal{F}}^{\dagger}\left[\frac{\hat{p}^{2}}{\kappa M(\tau)}+\hat{q}^{2}\right]\hat{\mathcal{F}}|\Psi(\tau)\rangle
\end{equation}
By noting that $e^{-i\frac{\theta}{2}(\hat{p}^{2}+\hat{q}^{2})}\hat{q}e^{i\frac{\theta}{2}(\hat{p}^{2}+\hat{q}^{2})}=\hat{q}\cos\theta-\hat{p}\sin\theta$
and $e^{-i\frac{\theta}{2}(\hat{p}^{2}+\hat{q}^{2})}\hat{p}e^{i\frac{\theta}{2}(\hat{p}^{2}+\hat{q}^{2})}=\hat{p}\cos\theta+\hat{q}\sin\theta$
we obtain
\begin{equation}
i\frac{\partial|\Psi(\tau)\rangle}{\partial\tau}=\frac{1}{2}\left[\hat{p}^{2}+\frac{\hat{q}^{2}}{\kappa M(\tau)}\right]|\Psi(\tau)\rangle\label{massdep}
\end{equation}

\section{Invariants and solutions}

It is well-known that the above equation has an invariant of the form
\cite{Lewis}
\begin{equation}
\hat{I}=\frac{1}{2}\left(\frac{\hat{q}^{2}}{\rho^{2}}+(\rho\hat{p}-\dot{\rho}\hat{q})^{2}\right)
\end{equation}
where the auxiliary function $\rho$, obeys the Ermakov equation
\begin{equation}
\ddot{\rho}+\frac{\rho}{\kappa M(\tau)}=\frac{1}{\rho^{3}}
\end{equation}
and is related to another auxiliary function $u$ by the equations
\begin{equation}
u=\rho\cos\left(\int\frac{d\tau}{\rho^{2}}\right),\qquad\rho=u\sqrt{1+\left(\int\frac{d\tau}{u^{2}}\right)^{2}}\label{urho}
\end{equation}
where $u$ obeys the equation
\begin{equation}
\ddot{u}+\frac{u}{\kappa M(\tau)}=0.\label{tdho}
\end{equation}
By using the unitary operator $\hat{T}$
\begin{equation}
\hat{T}=e^{i\frac{
\ln\rho}{2}(\hat{q}\hat{p}+\hat{p}\hat{q}+
\frac{2\rho\dot{\rho}}{\rho^2-1}\hat{q}^{2})}.
\end{equation}
that may be rewritten as (see the Appendix)
\begin{equation}
\hat{T}=e^{i\frac{\ln\rho}{2}(\hat{q}\hat{p}+\hat{p}\hat{q})}e^{-i\frac{\dot{\rho}}{2\rho}\hat{q}^{2}},\label{transfor}
\end{equation}
we can relate the invariant, $\hat{I}$, to the time independent harmonic
oscillator Hamiltonian,
\begin{equation}
\hat{T}\hat{I}\hat{T}^{\dagger}=\frac{1}{2}(\hat{p}^{2}+\hat{q}^{2}).\label{trai}
\end{equation}
This relation hints us that the transformation operator
(\ref{transfor}) may be relevant in the solution of
(\ref{massdep}). Indeed, it has been shown that the Schr\"odinger
equation for the time dependent harmonic Hamiltonian has a
solution of the form \cite{Moya03}
\begin{equation}
|\Psi(t)\rangle=e^{-i\hat{I}\int_{0}^{\tau}\omega(t){dt}}\hat{T}^{\dagger}\hat{T}(0)|\Psi(0)\rangle.\label{solut}
\end{equation}
with the characteristic frequency of the time dependent harmonic oscillator
defined by \cite{Guasti2} $\omega(t)=1/\rho^{2}$.

\subsection{Hyperbolically growing mass}
\begin{figure}[ht]
\centering{}\includegraphics[width=12cm]{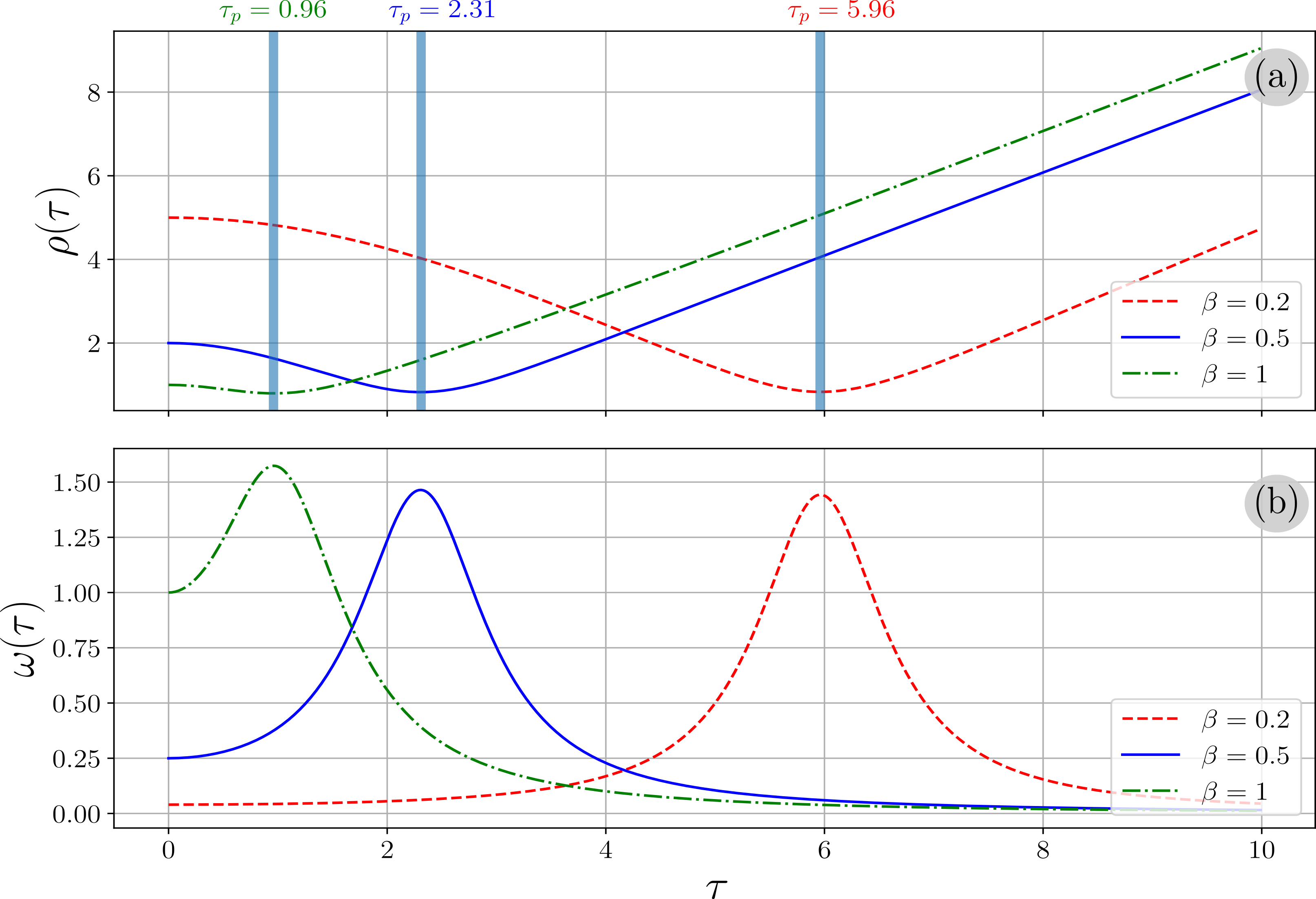}
\caption{\label{fig1} We plot the functions (a) $\rho(\tau)$ and
(b) $\omega(\tau)$ for $\beta=0.2$ (dot-dashed line), $\beta=0.5$
(dashed line) and $\beta=1$ (solid line).}
\end{figure}

We consider now a time dependent mass of the form $M(\tau)=\frac{\cosh^{2}(\beta\tau)}{2\beta^{2}\kappa}$.
For this choice we obtain the solution $u=\tanh(\beta\tau)$ and therefore
the auxiliary function, $\rho$, is given by
\begin{equation}
\rho=\tanh{\beta\tau}\sqrt{1+\frac{[\beta\tau-\coth(\beta\tau)]^{2}}{\beta^{2}}},
\end{equation}
and the integral in one of the exponentials of the wavefunction in
equation (\ref{cohevol}) may be calculated by the relation
\begin{equation}
\int\omega(\tau)d\tau=\cos^{-1}\frac{1}{\sqrt{1+\frac{[\beta\tau-\coth(\beta\tau)]^{2}}{\beta^{2}}}}
\end{equation}

If we consider the initial state to be a coherent state
\begin{equation}
|\psi(0)\rangle=|\alpha\rangle=e^{-\frac{|\alpha|^{2}}{2}}\sum_{n=0}^{\infty}\frac{\alpha^{n}}{\sqrt{n!}}|n\rangle=\hat{D}(\alpha)|0\rangle\label{ini}
\end{equation}
where $\hat{D}(\alpha)$ is the so-called Glauber displacement operator
\cite{Glauber}
\begin{equation}
\hat{D}(\alpha)=e^{\alpha\hat{a}^{\dagger}-\alpha^{*}\hat{a}},\label{dis}
\end{equation}
with the annihilation and creation operators given by
\begin{equation}
\hat{a}=\frac{1}{\sqrt{2}}(\hat{q}+i\hat{p}),\qquad\hat{a}^{\dagger}=\frac{1}{\sqrt{2}}(\hat{q}-i\hat{p}),\label{anni}
\end{equation}
and where the states $|n\rangle$ are Fock states. Then the Fourier
transformed initial wavefunction becomes $|\Psi(0)\rangle=|-i\alpha\rangle$
and so the evolved state has the form
\begin{equation}
|\Psi(\tau)\rangle=\hat{T}^{\dagger}|-i\alpha e^{-i\int_{0}^{\tau}\omega(t)dt}\rangle,\label{cohevol}
\end{equation}
or, more explicitely
\begin{equation}
|\Psi(\tau)\rangle=e^{i\frac{\dot{\rho}}{2\rho}\hat{q}^{2}}e^{-i\frac{\ln\rho}{2}(\hat{q}\hat{p}+\hat{p}\hat{q})}|-i\alpha e^{-i\int_{0}^{\tau}\omega(t)dt}\rangle,\label{cohevol2}
\end{equation}
this is the application of squeeze operators to a coherent state,
i.e, a squeezed state \cite{Yuen,Caves,Satya,Vidiella,Knight}. For
this particular example, figure 1 (a) shows that there is an specific
time, $\tau_{p}$, for which the derivative of the function $\rho(\tau)$
has a minimum, and therefore $\dot{\rho}(\tau_{p})=0$, for which
we obtain the exact squeezed state
\begin{equation}
|\Psi(\tau_{p})\rangle=|-i\alpha
e^{-i\int_{0}^{\tau}\omega(t)dt},\ln\rho(\tau_{p})\rangle,\label{cohevol2}
\end{equation}
and therefore, going back to the original picture, by transforming
this state with the inverse Fourier operator we obtain finally
\begin{equation}
|\psi(\tau_{p})\rangle=|i\alpha e^{-i\int_{0}^{\tau}\omega(t)dt},\ln\rho(\tau_{p})\rangle.\label{cohevol3}
\end{equation}
In figure 1 we also show a plot of the characteristic frequency of
the time dependent harmonic oscillator.

\begin{figure}[ht]
\centering{}\includegraphics[width=12cm]{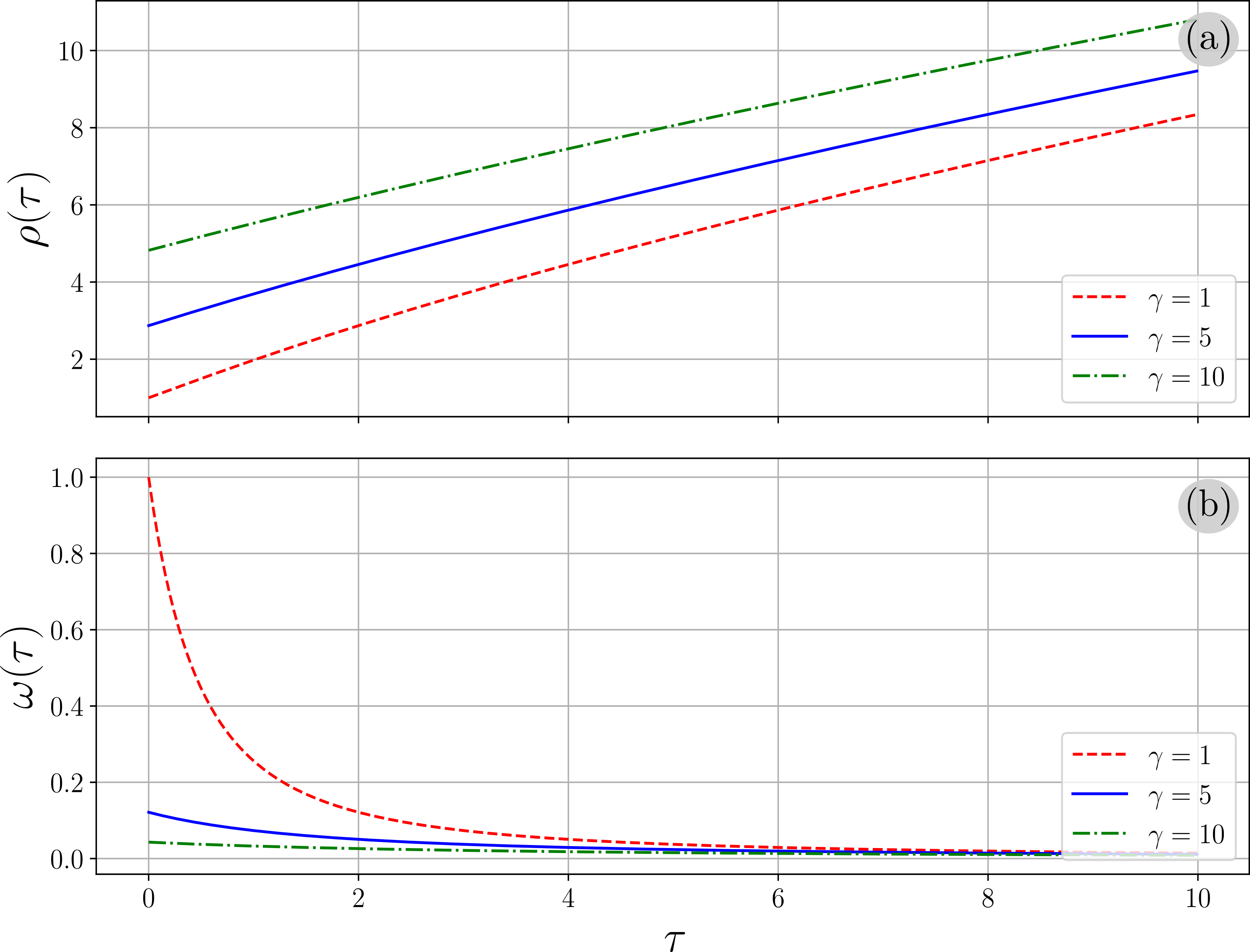}
\caption{\label{fig2}We plot the functions (a) $\rho(\tau)$ and
(b) $\omega(\tau)$ for $\gamma=1$ (dot-dashed line), $\gamma=5$
(dashed line) and $\gamma=10$ (solid line).}
\end{figure}

\subsection{ Quadratically growing mass}

We consider now a mass that evolves as
\begin{equation}
\kappa M(\tau)=(\gamma+2\tau)^{2}
\end{equation}
the solution to (\ref{tdho}) reads then
\begin{equation}
u(\tau)=\sqrt{\gamma+2\tau}
\end{equation}
and therefore the auxiliary Ermakov function reads
\begin{equation}
\rho(\tau)=\sqrt{(\gamma+{2}\tau)\left[1+\frac{1}{4}\ln^{2}(\gamma+{2}\tau)\right]}
\end{equation}
in this case, the integral in on of the exponentials of equation
(\ref{cohevol}) is given by
\begin{equation}
\int\omega(\tau)d\tau=\cos^{-1}\frac{1}{\sqrt{1+\frac{1}{4}\ln^{2}(\gamma+{2}\tau)}}
\end{equation}
In figure 2 we show the behaviour of (a) the auxiliary function $\rho(\tau)$
and (b) the characteristic frequency of the time dependent harmonic
oscillator. In this case there is no way to simplify the state to
an ideal squeezed state as the function $\rho(\tau)$ does not show either maximums nor minimums.

\section{Conclusions}
It has been shown that if the Fourier operator is used to
transform the quantum time dependent harmonic oscillator
Hamiltonian when a time dependent mass is considered, the
resultant Hamiltonian takes the form of one with time dependent
frequency. By writing the invariant for this latter Hamiltonian,
we can generate solutions that involve squeeze operators. We have
given two examples of time dependent mass, namely hyperbolically
and quadratically growing masses. Finally, by using the relation
(\ref{urho}), we have been able to show how to calculate the
integral of the characteristic time dependent frequency needed in
(\ref{cohevol}).
\section*{Appendix}
The exponential
\begin{equation}
e^{i\frac{\ln\rho}{2}(\hat{q}\hat{p}+\hat{p}\hat{q}+
\frac{2\rho\dot{\rho}}{1-\rho^2}\hat{q}^{2})}
\end{equation}
has the sum of the operators
\begin{equation}
\hat{A}=i\frac{\ln\rho}{2}(\hat{q}\hat{p}+\hat{p}\hat{q})
\end{equation}
and
\begin{equation}
\hat{B}=i \ln\rho \frac{\rho\dot{\rho}}{\rho^2-1}\hat{q}^{2}.
\end{equation}
We can show that the commutators of the above operators is
\begin{equation}
[\hat{A},\hat{B}] = 2 \ln\rho\hat{B}.
\end{equation}
Using the fact that the commutator of the two operators is
proportional to one of them, we can factor the exponential in the form
\begin{equation}
e^{i\frac{\ln\rho}{2}(\hat{q}\hat{p}+\hat{p}\hat{q}+
\frac{2\rho\dot{\rho}}{1-\rho^2}\hat{q}^{2})} = e^{i\frac{\ln\rho}{2}(\hat{q}\hat{p}+\hat{p}\hat{q}%
)}e^{-i\frac{\dot{\rho}}{2\rho}\hat{q}^{2}}
\end{equation}\bigskip{}
\bigskip{}

\end{document}